\documentclass[11pt,english,a4paper]{article}
\pdfoutput=1
\usepackage{amsmath}
\usepackage{amssymb}
\usepackage{graphicx}
\usepackage{epsfig,jheppub}
\usepackage{float}
\usepackage{hyperref}
\usepackage{multirow}
\usepackage{subfig}
\makeatletter

\title{Thermodynamic Properties of Static and Rotating Unparticle Black Holes}

\author[a]{G. Alencar,}
\author[b]{C. R. Muniz}
\affiliation[a]{Departamento de F\'{\i}sica, Universidade Federal do Cear\'{a},
Caixa Postal 6030, Campus do Pici, CEP 60455-760, Fortaleza, Cear\'{a}, Brazil.\\}
\affiliation[b]{Universidade Estadual do Cear\'a, Faculdade de Educa\c c\~ao, Ci\^encias e Letras de Iguatu-
Av. D\'ario Rab\^elo, s/n - Vila Santo Ant\^onio, Iguatu - CE, CEP: 63502-253, Brazil.
}

\emailAdd{geova@fisica.ufc.br}
\emailAdd{celio.muniz@uece.br}

\abstract{
In this paper we find analytical expressions for thermodynamic quantities of scalar (tensor) and vector unparticle static black holes. We also find rotating solutions to these systems and analyse their thermodynamics. First we consider the static case with a spherically symmetric source for both the vector and scalar (tensor) unparticles. We obtain thus analytical expressions  to the principal thermodynamic quantities: Hawking temperature, entropy, heat capacity and free energy. For the scalar (tensor) case we find that the black hole presents a residual value for the entropy when its radius goes to zero but the other thermodynamic quantities give, for any horizon radius, a thermodynamically unstable behavior similar to the standard black hole. For the vector case we find a richer structure in the region in which the horizon radius is less than the characteristic length of the unparticle theory. We identify a phase transition and a region where the black hole can be thermodynamically stable. Following, we show that the mentioned modifications in the standard gravity are formally similar to those ones present in the black holes with quintessence. With this we also show, notwithstanding, that the unparticles cannot be a source of quintessence. By using this similarity we find two different rotating solutions to the unparticle black holes based on works by Ghosh and Toshmatov {\it et al}. For both cases we compute the Hawking temperature and in the ungravity dominated regime we find, as in the static cases, a fractalization of the event horizon.  For the Gosh-like solution the fractal dimension depends on the polar angle and on the rotation of the source. For the Toshmatov-like one it is equal to the static case and therefore the fractalization is not dependent on the rotation of the source.}


\begin{document}
\maketitle
\section{Introduction}
Various types of fundamental physical symmetries are present in the Standard Model (SM) of elementary particles and fields. However there is one that plays relevant roles in several areas of contemporary theoretical physics and which is left out, at least at the energy scales currently accessible to the mankind: that one associated with scale transformations in both distance and time. Thus, for instance, in condensed matter physics, phase transitions can occur near a critical point and such a feature needs a scale invariant description \cite{ZinnJustin:2002ru}. In quantum and classical theories of massless fields, the scale invariance points to the nonexistence of a fundamental length scale \cite{Polchinski:1987dy}. By the other hand, in the context of quantum gravity theories, which admit such a length, as the Horava-Lifshitz one, the breaking of that invariance from establishment of different scale transformations for space and time around a critical point in ultraviolet regimes violates the local Lorentz symmetry and guarantees renormalization by power counting \cite{Horava:2009uw,Horava:2008ih,Horava:2009if,Visser:2011mf}. It is worth mentioning that the study of black holes in this scenario of scale invariance breakdown shed some light on various features of the gravity behaviour at very high energies \cite{Kehagias:2009is,Muniz:2013uva,Muniz:2014dga,Alencar:2015aea}.

A question that naturally arises is if it is possible to have scale invariance in an infrared sector of SM, since in this regime the particles possesses mass and such a symmetry in principle cannot exist. Thus, a proposal inspired in the old Bank-Zaks (BZ) theory \cite{Banks:1981nn} was made ten years ago, in which a conformal invariant high energy sector near a critical (fixed) point is possible, where there would be fields of unknown nature which would be very weakly coupled to those ones of the standard model, since the coupling constant depends on the inverse of the mass associated with that very high energy scale. In fact, the corresponding interaction Lagrangian has the form \cite{Georgi:2007ek,Georgi:2007si}
\begin{equation}
\mathcal{L}=\frac{1}{M^k_U}\mathcal{O}_{SM}\mathcal{O}_{BZ},
\end{equation}
where $k=d_{SM}+d_{BZ}-4$, with $d_{SM}$ and $d_{BZ}$ being the scale dimensions of the SM and BZ field operators, respectively. Furthermore, around some lower energy scale $\Lambda_U<M_U$ (one hopes that it be of the order of some TeV's) a dimensional transmutation in the hidden sector field operator, $\mathcal{O}_{BZ}\rightarrow, \mathcal{O}_{U}$, would occur, permitting stronger interactions between the novel entities which arise - the so called {\it unparticles} - and the particles of the SM, with phenomenological implications accessible in principle. This transmutation would yield scale-invariant interaction terms with fractionary scale dimension $d_U$, implying a non-trivial phase space structure. Since then the unparticle idea has produced a substantial quantity of research by considering modifications to the known physics, beyond the SM, as in high energy particle phenomenology \cite{Luo:2007bq, Aliev:2007qw,Chen:2007qr, Mathews:2007hr,Bander:2007nd, Cheung:2007ap, Kikuchi:2007qd, Zhang:2007ih, Cheung:2007jb} and in astrophysical and cosmological scenarios, including those ones with black holes \cite{Lewis:2007ss,McDonald:2007bt,Davoudiasl:2007jr,Das:2007nu,Hannestad:2007ys,Alberghi:2007vc,Gaete:2010sp,Mureika:2010je,Mureika:2007nc}. In these latter one draws the attention to the fact that the exterior event horizon presents a surface with fractal dimension equals to $d_U$. It is noticeable also that a recent paper on Casimir effect with unparticles shows a fractalization in the parallel plates dimension \cite{Frassino:2013lya}.

Regarding unparticle black holes, in the above cited papers the rich thermodynamical properties of these exotic objects were studied by means of an incomplete approach, since only the extreme regimes of un-gravity and gravity dominated phases were separately analysed. In the present work, we will go to a deeper level of understanding and those properties will be analytically obtained by considering both the vector and scalar (tensor) unparticle modifications of the Einstein's gravity of a spherically symmetric source, for all the ranges. Features as phase transitions will be also investigated and comparisons with the known characteristics of the usual black holes will be made. We will show that the mentioned modifications in the standard gravity are formally similar to those ones present in black holes with quintessence \cite{Kiselev:2002dx}, although the unparticles themselves cannot be source of quintessence, as it will become clear. Moreover, this mathematical mapping of the unparticle black holes into the quintessential ones will allow us to build out rotating solutions to those ones, about which we will find some thermodynamics properties too.

This paper is organized as follows: In section two we review the black hole solutions for scalar (tensor) and vector unparticles. We also review the thermodynamic properties of these black holes in the asymptotic regimes. In section three we construct analytical solutions to the mass parameter, temperature, entropy, heat capacity and free energy for theses black holes in order to study their thermodynamic properties for all ranges. Finally in section four we construct the solution for the rotating unparticle black holes.

\section{A Review of Unparticle Static Black Holes}

In this section we must review the solution to the scalar (tensor) and vector unparticle static black holes found in Refs.  \cite{Gaete:2010sp,Mureika:2010je}. The action for the system is given by $S=S_M+S_U$, where $S_M$ is the matter action
\begin{equation}
 S_M\equiv -\int d^4x \sqrt{g}\,\rho\left(\, x\,\right)\, u^\mu\, u^\nu\ ,\quad
 \rho\left(\, x\,\right)\equiv
 \frac{M}{\sqrt{g}}\int d\tau \,\delta\left(\, x -x\left(\tau\right)\,\right)
\end{equation}
and $S_U$ is the sum of the Einstein-Hilbert action and a correction due to unparticles
\begin{equation}
S_U = \frac{1}{2\kappa^2}\,\int d^4x \sqrt{g}\,\left[ \,
1+\frac{ A_{d_U}}{\left(\,2d_{U}-1\,\right)\sin\left(\,\pi\, d_U\,\right)}
\frac{\kappa_\ast^2}{\kappa^2}
\left(\, \frac{-D^2}{\Lambda^2_U}\,\right)^{1-d_U}\, \right]^{-1} R .
\label{ueh}
\end{equation}
In the above expression $D^2$ stands for the D'Lambertian,
\begin{equation}
A_{d_U}\equiv \frac{16\pi^{5/2}}{\left(\, 2\pi\,\right)^{2d_U}}
\frac{\Gamma\left(\,d_U + 1/2\,\right)}{ \Gamma\left(\,d_U - 1\,\right)
\Gamma\left(\,2d_U \,\right)}
 \end{equation}
and
\begin{equation}
\kappa_\ast \equiv \frac{1}{\Lambda_U}\left(\,
\frac{\Lambda_U}{M_U}\,\right)^{d_{UV}}.
\end{equation}
The strength of the coupling constant is determined by the mass scale $M_U$
which replaces the Planck mass. By assuming a static source the authors found the analytical solution
for the spherically symmetric line elements. They are given by

\begin{equation}\label{staticsolution}
g_{rr}^{-1}=-g_{00}=1-\frac{2M}{r}\left[\,1\pm\left(\,\frac{R_{s,v}}{r}\,\right)^{2d_{U}-2}\,\right]
\end{equation}
where
\[
R_{s,v}=\left[\,\Gamma_U\frac{M_{Pl.}^{2}\kappa_{s,v}^{2}}{\pi^{2d_{U}-1}}\,\Lambda_{U}^{2-2d_{U}}\right]^{\frac{1}{2d_{U}-2}} ;\;\Gamma_U=\frac{\Gamma\left(\,d_{U}-1/2\,\right)\Gamma\left(\,d_{U}+1/2\,\right)}{\Gamma\left(\,2d_{U}\,\right)}\, ,
\]
$k_{s,v}$ are the coupling of the $s,v$ unparticles with gravity and $M_{Pl}$ is the plank mass.  The plus signal is taken for the scalar (tensor) unparticle case ($R_{s}$) and
the minus one for the vector case ($R_{v}$). These
solutions has horizon curves defined by $g_{rr}^{-1}=0$
\begin{equation}\label{statichorizon}
 M=\frac{r_H}{2}\frac{1}{1 \pm\left(\,R_{s,v} /r_H\,\right)^{2d_U-2}}.
\end{equation}

The authors in Refs \cite{Gaete:2010sp,Mureika:2010je} distinguish between two
regimes. The gravity dominated (GD) regime  is defined by $R_{s,v}\ll r_{H}$ and  the ungravity dominated (UGD) regime  by $r_{H}\ll R_{s,v}$. In the GD regime, for both cases,  the mass parameter reduces to standard $M=r_H/2$. However in the UGD regime
the mass is given by
\begin{equation}\label{massUGDR}
 M\approx\pm\frac{1}{2}\frac{r_H^{2d_U-1}}{R_{s,v}^{2d_U-2}}.
\end{equation}

Therefore, for the vector case, the mass parameter becomes negative in the UGD regime. In fact for the entire region $r_H<R_v$ this parameter is negative. This fact lead the
authors of Ref. \cite{Mureika:2010je} to ignore this region. However it is a known fact that a negative mass parameter indicates black hole solution with nontrivial topology \cite{Mann:1997jb}. This suggests that beyond the fractalization of the horizon the vector unparticle
can in fact contribute to changes in the black hole topology. Since the UGDR regime must be contained in a full quantum gravity theory, it is expected that
quantum fluctuations in geometry should enhance quantum fluctuations in topology \cite{Wheeler:1957mu}. In fact in the incoming sections we will see that this is a very rich region, with phase transitions and stable black holes. The Hawking temperature can also be computed. It is given by \cite{Wald:1984rg}
\begin{equation}\label{tbh}
T_H=\frac{\kappa}{2\pi},
\end{equation}
where $\kappa$ is the surface gravity which in this case is given $\kappa=g'_{00}(r_H)/2$ and we get
\begin{equation}\label{BHT}
 T_{d_U}=
 \frac{1}{ 4\pi\, r_H\left[\, 1 \pm\left(\, \frac{R_{s,v}}{r_H} \,\right)^{2d_U-2} \,\right]  }
 \left[\, 1 \pm\left(\, 2d_u-1\,\right)\,\left(\, \frac{R_{s,v}}{r_H} \,\right)^{2d_U-2}\,\right].
\end{equation}

 In the GD regime the temperature reduces to the
usual expression for the Hawking temperature. However in the UGD regime
it is given by
\begin{equation}\label{bhtUGDR}
T_{d_U}\approx \frac{2d_U-1}{4\pi r_H}.
\end{equation}
By comparing this with the Hawking temperature for a Schwarzschild black hole in $D$ spacetime dimensions \cite{Emparan:2008eg}
\begin{equation}
T_{D}\approx \frac{D-3}{4\pi r_H}
\end{equation}
and than they argue that unparticle gravity leads to a fractalization of the event horizon with
\begin{equation} \label{fractalization}
d_H=2d_U; d_H\equiv D-2.
\end{equation}
In order to enforce the fractalization of the event horizon in the UGD regime the authors compute the entropy of the system. At this point we must be careful. It is very easy to see that the standard expression for entropy $S=\pi r^2_{H}$ is not correct for this corresponds to $dS=2\pi r$ which is very different of $dM/T$ as given above. Therefore the correct procedure is to obtain the entropy by integrating $dS=dM/T$.
In the UGD regime we use Eqs. (\ref{bhtUGDR}) and (\ref{massUGDR}) to obtain
\begin{equation}
dS=\pm2\pi\frac{r_H^{2d_U-1}}{R_{s,v}^{2d_U-2}}dr_H,
\end{equation}
and
\begin{equation}\label{UGDentropy}
S=\pm\frac{\pi}{d_U}\frac{r_H^{2d_U}}{R_{s,v}^{2d_U-2}}.
\end{equation}
Some points are worth noticing about this result. First we see that the entropy in fact depend on an effective dimension given by $d_H=2d_U$. Second, for the vector case it seems that, beyond having a negative mass, at this region we have negative entropy. In the next section we must treat this problem in detail by finding an analytical solution to the entropy and free energy of the system. We can advance that our result show that a constant must be added in above entropy, rendering it positive even in the vector case.

\section{Analytical Thermodynamic Properties of Static Unparticle Black Holes}

As said in the last section the approximate analysis of the black hole properties for the gravity ($r_H\gg R_{s,v}$) and un-gravity ($r_H\ll R_{s,v}$) dominated regimes has been done in Refs \cite{Gaete:2010sp,Mureika:2010je}. Here we extend the analysis and consider intermediate values of $r_{H}$. For this we first analyze carefully the mass an temperature. After we find analytical solutions for entropy and free energy and finally we
discuss the possibility that unparticles can describe quintessence.

\subsection{Mass Parameter and Temperature}

In this subsection we analyze the mass parameter an temperature for the scalar (tensor) and vector unparticle black holes. First of all we should point that for the unvector case, with minus sign, Eq. (\ref{statichorizon}) has some peculiarities. As said before the mass parameter is always negative (positive) if $r_H<R_v$ ($r_H>R_v$) and diverges if $r_H=R_v$. This suggests that we have two disjointed regions , {\it i.e.}, we cannot transit
from one to the other. Despite the negativeness of the mass parameter $M$ recent works points to the fact black holes with $M<0$ can exist and have non-trivial topology \cite{Mann:1997jb}. Despite of the richness of this subject here we focus on the analysis of the thermodynamic properties of this kind of black holes. We will see that in some regions they are well defined and, more than this, we can have thermodynamically stable black holes.

\begin{figure}[!ht]
     \subfloat[$R_{s,v}=1$ and $0<r_{H}<1.5$\label{TempUngravityRange1}]{%
      \includegraphics[width=0.5\textwidth]{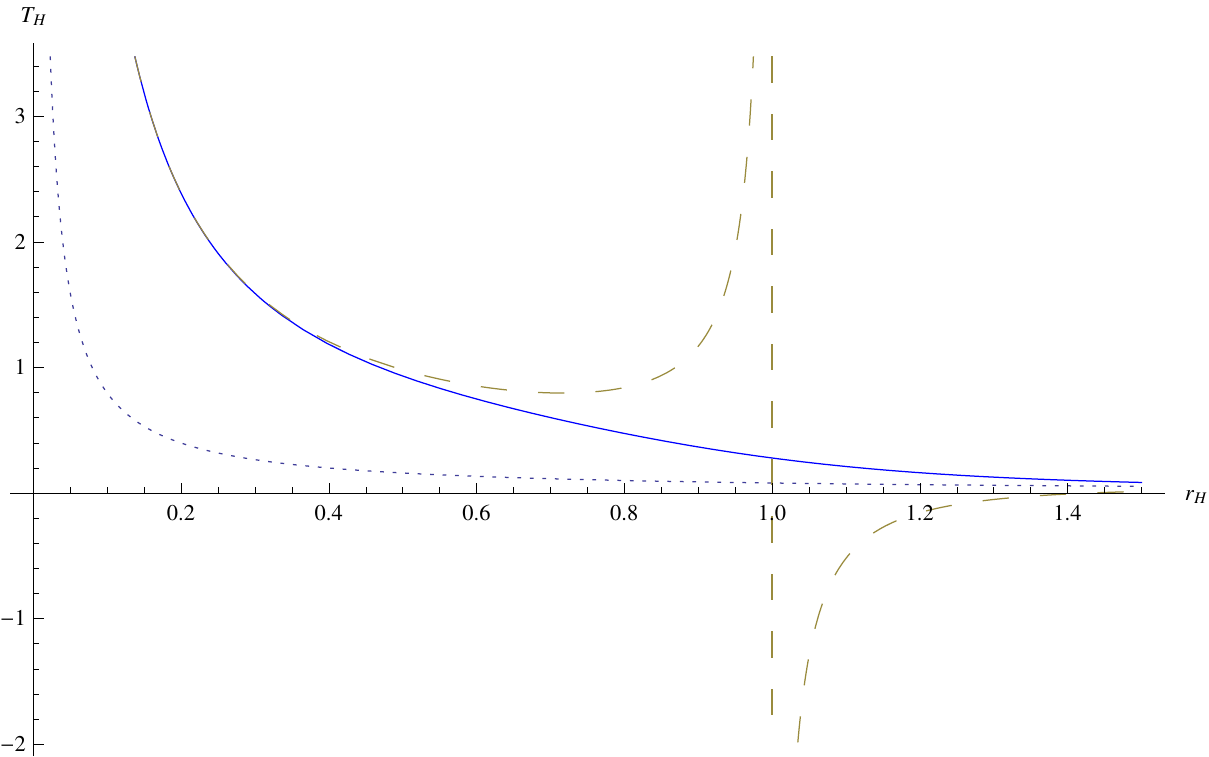}
     }
     \hfill
     \subfloat[$R_{s,v}=1$ and $1.5<r_{H}<4$\label{TempUngravityRange2}]{%
       \includegraphics[width=0.5\textwidth]{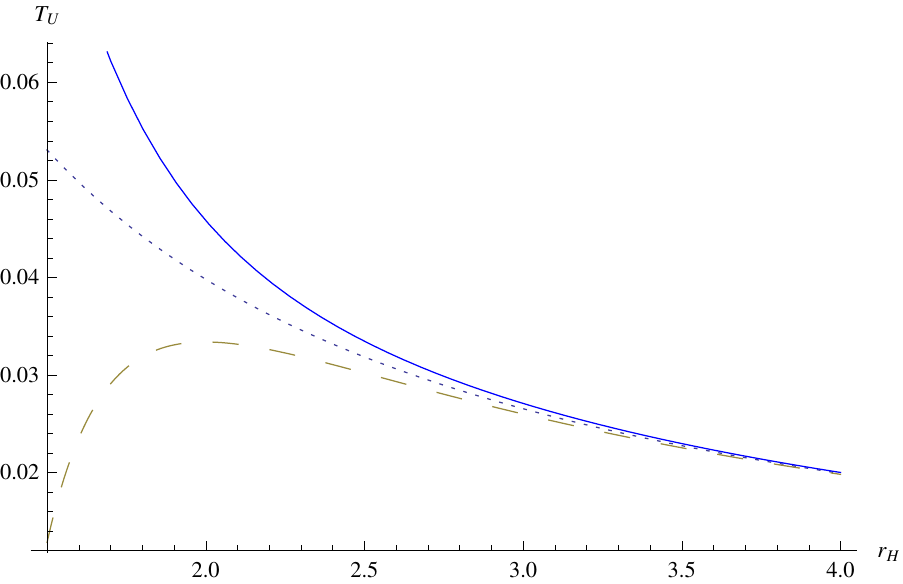}
     }
     \caption{The BH temperature for the standard, vector and scalar (tensor) unparticle cases depicted in dotted, dashed and solid lines respectively. }
     \label{TempUngravity}
   \end{figure}

The temperature  for the standard, scalar (tensor) and vector cases is depicted in Figs. \ref{TempUngravityRange1} and \ref{TempUngravityRange2}. Initially we can observe that in the range $r_H<R_{s,v}$ the standard black hole is colder than the scalar (tensor) and vector cases, but for the range $r_H>R_{s,v}$ it presents an intermediate temperature between the scalar (tensor) and vector unparticle black holes.  We also can see that in GD and UGD regimes the behavior of the temperatures for all cases are similar. However, for intermediate values of $r_H$, we have more involved behaviors of the temperature for the un-vector case.
In fact, in this case two stationary points for the temperature can be easily found and are given by
\begin{equation}\label{roots}
r_H^{\pm}=R_v\left(\frac{2-3d_U+2d^2_U\pm \sqrt{5-14d_U+17d^2_U-12d^3_U+4d^2_U}}{2d_U-1}  \right)^{\frac{1}{2-2d_U}}
\end{equation}
where $r_H^{-}$ ($r_H^{+}$) corresponds to the minimum (maximum) of the temperature. This suggests the presence of a phase transition in $r_H^{-}$. In the same figure we also see that in the region around $r_{H}\approx R_{v}$ the temperature is discontinuous, what reinforces the previously mentioned disjointedness of the regions $ r_H\lessgtr R_v $.  In Fig. \ref{TempUngravityRange2} we also see that we have a vanishing temperature. In order to discover if the point $r_H^{-}$ in fact corresponds to some phase transition we must find the heat capacity and the free energy of the system. This is the topic of the next subsection.

\subsection{Analytical Entropy, Heat Capacity and Free Energy}

Now we must focus in obtaining the an analytical expressions for the heat capacity and free energy in order to loog for phase transitions of the system. Before this we must analyze the next important variable of the system, which is the entropy. This will automatically provide us an analytical expression for the free energy. As pointed in the last sections, the correct procedure is to obtain the entropy directly from the definition
\begin{equation}\label{primeiraordem}
dS=\frac{M'(r_{H})}{T_{d_{U}}}dr_{H}=\frac{2\pi r_H}{1\pm\left(\frac{R_{s,v}}{r_H}\right)^{2d_U-2}}dr_{H},
\end{equation}
where $M$ and $T_{d_U}$ are given by Eqs. (\ref{statichorizon}) and (\ref{BHT}). For small values of $R_{s,v}$ the solution of the above equation must recover the standard area law, which is given by $\pi r_{H}^{2}$. With this we try the anzats $S=\pi r_{H}^{2}\Omega(r_H(z))$ with $z=(R_{s,v}/r_H)^{2d_U-2}$ and get the equation
\begin{equation}
z(1\pm z)\Omega '(z)+ \frac{(1\pm z)}{1-d_U}\Omega -\frac{1}{1-d_U}=0.
\end{equation}
The above equation has a very similar structure as the Euler's hypergeometric differential equation (EHDE)
\begin{equation}
z(1- z)\omega''+\left[c- z\left(1+a+b\right)\right]\omega'- ab\omega=0.
\end{equation}
 In fact if we differentiate  it we get
\begin{equation}\label{entropyequation}
z(1\pm z)\Omega''+\left[1+\frac{1}{1-d_U}\pm z\left(2+\frac{1}{1-d_U} \right)\right]\Omega'\pm\frac{1}{1-d_U}\Omega=0
\end{equation}
where the plus(minus) signal stand for the un-scalar(vector) case. The equations are the same if we identify
\begin{equation}
a=1,b=1/(1-d_U), c=(2-d_U)/(1-d_U)
\end{equation}
and for the scalar (tensor) case by performing $z\to-z$. With this we get that, to find solutions of equation (\ref{primeiraordem}), we can first look for solution to the EHDE. However it is known that  the solutions to the EHDE are build out from the hypergeometric series $_2F_1(a,b;c;\pm z)$.

First we consider the scalar (tensor) case in details, since it contains some subtleties  not considered previously by the authors of Ref. \cite{Gaete:2010sp}.  We have that the above equation has no singular points since $z>0$. In order to find the solution we can perform the change $z\to-z$.  When $c$ in not integer, the solution around $z=0$ is given by
\begin{equation}
\Omega(z)=A_s z^{1-c}\,_2F_1 (1+a-c,1+b-c;2-c;-z)+B_s \,_2F_1 (a,b;c;- z)
\end{equation}
where $A_s,B_s$ are constants of integration. In our case we have $1+b-c=0$ and therefore our solution simplifies to
\begin{equation}
\Omega(z)=A_s z^{\frac{1}{d_U-1}}+B_s \,_2F_1 (a,b;c;-z)
\end{equation}
and, as said above, this solution is well defined for all $z>0$. We should point that we are looking for the solution of the first order equation (\ref{primeiraordem}) and with this we get that one of the above constants will be fixed by substituting the above solution in (\ref{primeiraordem}). We finally obtain the entropy in terms of our original coordinate $r_{H}$
\begin{equation}\label{scalarentropy}
S_{s}=A_s+\pi r_H^2 \;_2F_1\left[1,\frac{1}{1-d_U},\frac{2-d_U}{1-d_U},- \left(\frac{R_{s}}{r_H}\right)^{2d_U-2} \right]
\end{equation}
where $A_s$ is an integration  constant. Since the above solution is valid for all $R_s$ the constant $A_s$ can be fixed as zero by demanding that in the limit $R_s\to 0$ the above expression recover the standard BH entropy.

For the vector case, equation (\ref{entropyequation}) has singular points and we must be careful. Differently of the scalar (tensor) case here we will have two diferent solutions in the regions $r_{H}\lessgtr R_{s,v}$. For the region $z<1$, since $c$ is not integer and $1+b-c=0$,  the solution around $z=0$ simplifies to
\begin{equation}\label{vectorentropy2}
S=\pi r_H^2 \;_2F_1\left[1,\frac{1}{1-d_U},\frac{2-d_U}{1-d_U}, \left(\frac{R_{v}}{r_H}\right)^{2d_U-2} \right]
\end{equation}
where again the integration constant has been fixed as zero in order to recover the standard entropy when $R_v\to 0$. As expected the above solution is singular when $z\to1$ ($r_{H}\to R_{v}$). For $z>1$ we have to expand our solution around $z=\infty$. This is possible since $a-b$ is not an integer and the solution to Eq. (\ref{entropyequation}) will be
\begin{equation}
\Omega(z)=A_v z^{\frac{1}{d_U-1}}- B_v z^{-1}\,_2F_1 \left(1,\frac{d_U}{d_U-1};\frac{1-2d_U}{1-d_U}; z^{-1}\right).
\end{equation}
Again, as we are looking for solutions of Eq. (\ref{primeiraordem}) the constant $B_{v}$ is fixed and we get the final expression for the entropy
\begin{equation}\label{vectorentropy1}
S_{v}=A_v-\pi r^2_{H}\left(\frac{r_H}{R_{v}}\right)^{2d_U-2}\,_2F_1 \left[1,\frac{d_U}{d_U-1};\frac{1-2d_U}{1-d_U}; \left(\frac{r_H}{R_{v}}\right)^{2d_U-2}\right].
\end{equation}
However in this region ($r_H<R_v$) the integration constant $A_v$ cannot be fixed as before by demanding that when $R_v/r_H\ll 1$ the standard entropy must be recovered. The only way of fixing this parameter is by analyzing phase transitions, as we will briefly see.

\begin{figure}[!ht]
\begin{center}
       \includegraphics[width=0.8\textwidth]{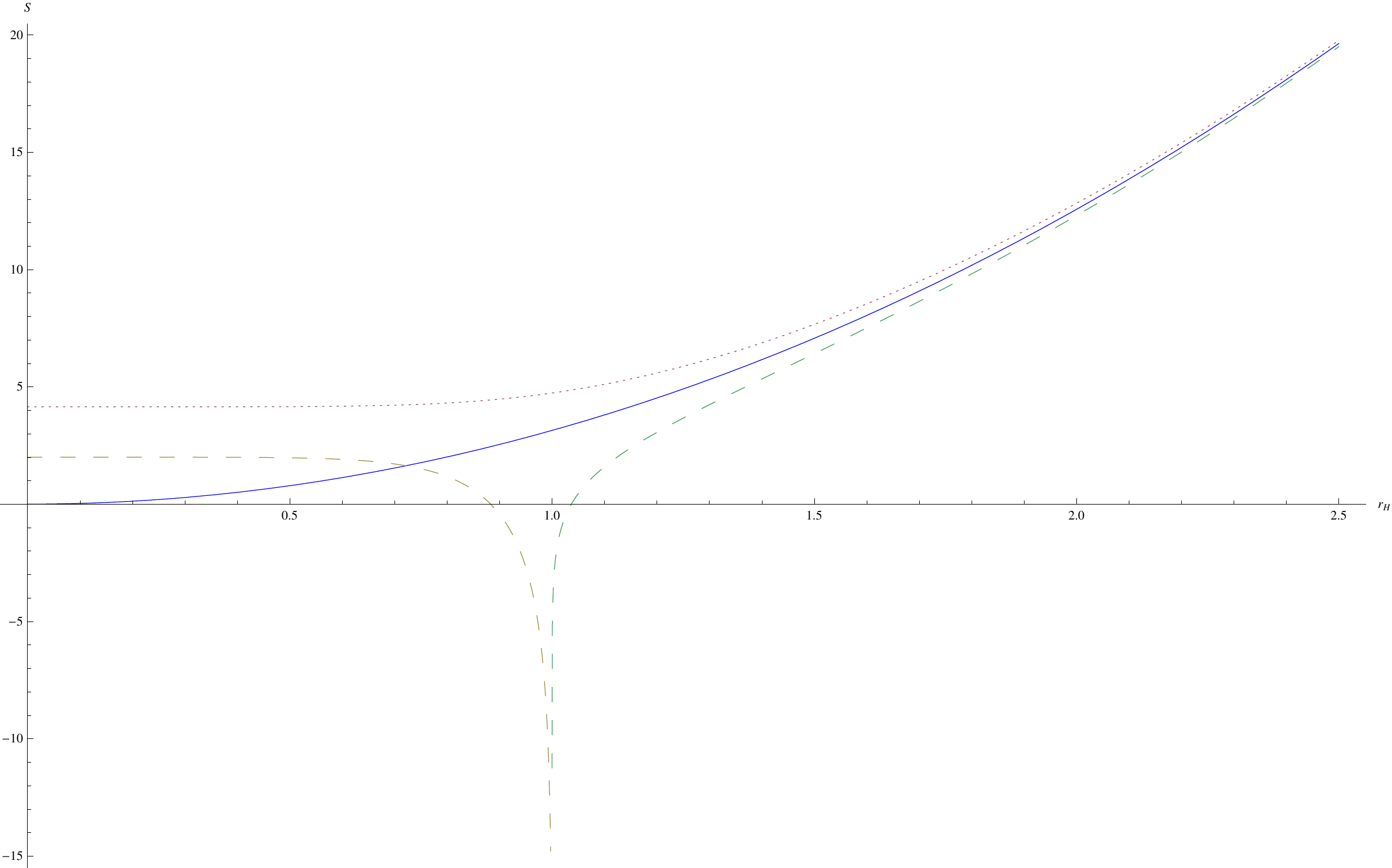}
    \caption{The entropy for the standard, scalar (tensor) and vector unparticles BH depicted in solid, dotted and dashed lines, respectively. This is given for the range $0<r_H<2$ and with $R_{s,v}=1$}
    \label{EntropyUngravity}
     \end{center}
   \end{figure}

In Fig. \ref{EntropyUngravity} we give the entropy for the standard, scalar (tensor) and vector unparticles BH in terms of $r_{H}$.
Despite the fact that the constant $A_s$ has been fixed to zero in the scalar (tensor) case, the above graph
show us that in the limit $r_H\to 0$ we get a constant value for the entropy. Therefore, as can be seen in Fig. \ref{EntropyUngravity} we arrive at the interesting result that, for different reasons, in both the scalar (tensor) and vector cases the entropy do not vanish when the black hole radius goes to zero. The fact the $A_v$ is not zero also guarantees that in some range we can get a positive entropy in the region $r_H<R_v$. We should point out here that the entropy given in Eq. (\ref{UGDentropy}), found in Ref. \cite{Gaete:2010sp}, for the scalar (tensor) and vector cases do not have this residual entropy. However their solution was found considering Eq. (\ref{primeiraordem}) in the region $r_H\ll R_{s}$ and they impose $S=0$. Our analytical solution shows that the only way to fit this is by considering $A_s\neq 0$. However this would imply that the standard area law is not recovered in the limit $R_{s}\to 0$ . Therefore our analysis show that the only way to obtain a full continuous solution that recover the standard black hole area law implies a finite value for the entropy in the limit $r_H\to 0$.

Finally we must study the phase transitions of the model. For this we need of the heat capacity and free energy of the system which are defined by
\begin{equation}
C_V=dM/dT;\; F=M-TS.
\end{equation}
The heat capacity can be obtained from Eqs. (\ref{statichorizon}) and (\ref{BHT})
\begin{equation}
C_V=\frac{M'}{T'_{d_U}}=-\frac{2 \pi  r_H^2  \left[1\pm (2 d_U-1) \left(\frac{R}{r_H}\right)^{2 d_U-2}\right]}{1 \pm2\left(2 d_U^2-3 d_U+2\right)  \left(\frac{R}{r_H}\right)^{2 d_U-2}+(2 d_U-1)  \left(\frac{R}{r_H}\right)^{4 d_U-4}}
\end{equation}
and the free energy for the vector case from Eqs. (\ref{statichorizon}), (\ref{BHT}), (\ref{vectorentropy2}) and (\ref{vectorentropy1}) .

\begin{figure}[!ht]
\begin{center}

       \includegraphics[width=0.7\textwidth]{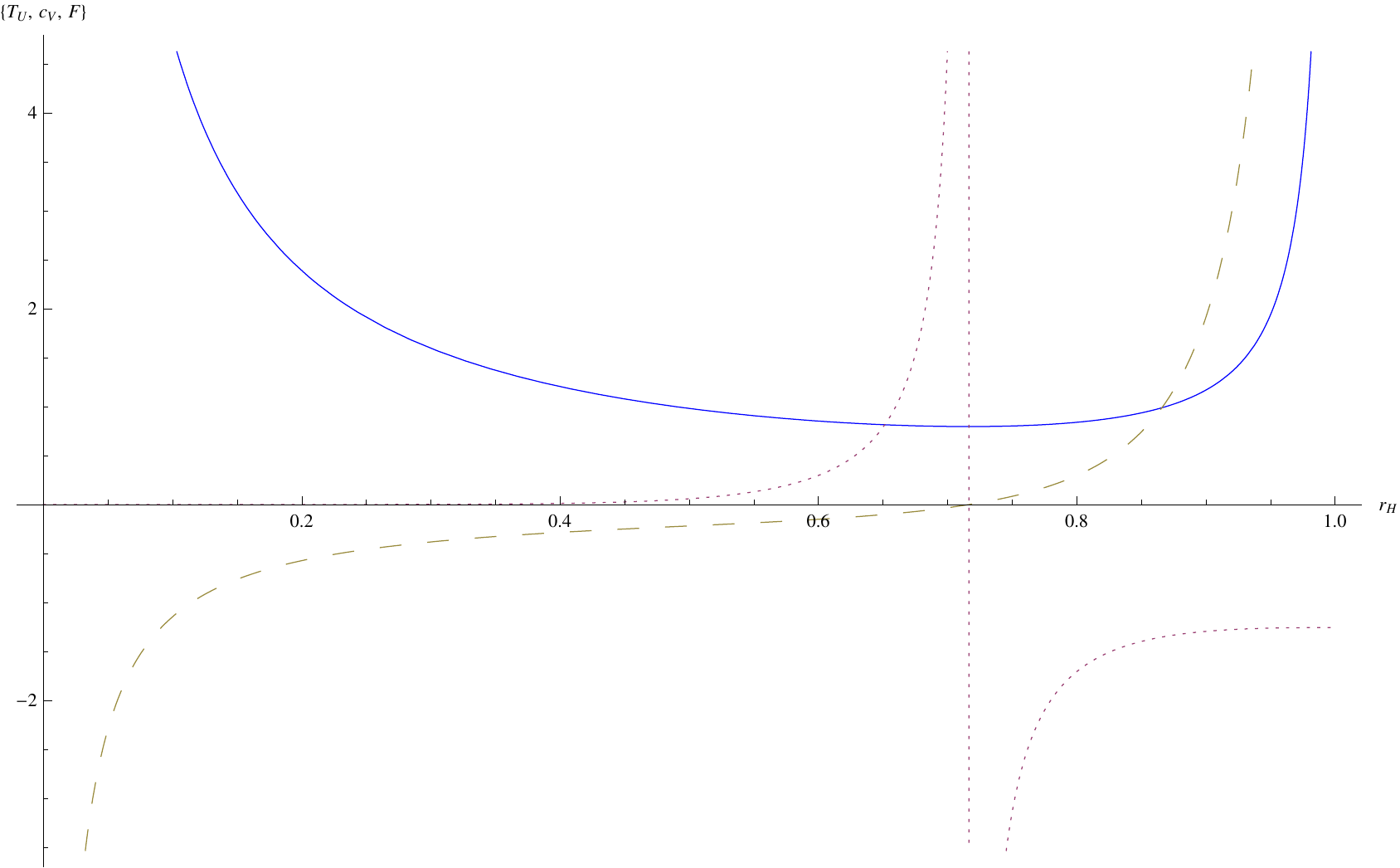}
     \caption{The vector unparticle BH  heat capacity, temperature and free energy depicted in dotted, solid and dashed lines, respectively. We consider the range $0<r_H<1$ and $R_v=1$}
     \label{Un-vectorThermoVariables}
    \end{center}
   \end{figure}

In Fig. \ref{Un-vectorThermoVariables} we depict the vector unparticle BH  heat capacity, temperature and free energy. We can see that the heat capacity confirms the fact that at the minimum of the temperature we have a phase transition. Therefore it seems natural to choose the value for $A_v$ such that the free energy reinforce such a phase transition. For this we have to impose the latter as being zero at the temperature minimum. Equation (\ref{roots}) gives the value of $r^{+}_H$ for the phase transition and by fixing that at the same point the free energy is zero we get
\begin{eqnarray}
&A_v&=\pi (r^{-}_H)^2\Bigg\{2\left[1-(2d_U-1)\left(\frac{R_v}{r^{-}_H}\right)^{2d_U-2}\right]+\nonumber\\
&&+\left(\frac{r^{-}_H}{R_v}\right)^{2d_U-2}\,_2F_1 \left[1,\frac{d_U}{d_U-1};\frac{1-2d_U}{1-d_U}; \left(\frac{r^-_H}{R_{v}}\right)^{2d_U-2}\right]\Bigg\}.
\end{eqnarray}

 \begin{figure}[!ht]
    \subfloat[$R_{v}=1$ and $1.1<r_{H}<2$\label{CapacityHeatTemperatureFreeEnergyVectorRange2}]{%
       \includegraphics[width=0.5\textwidth]{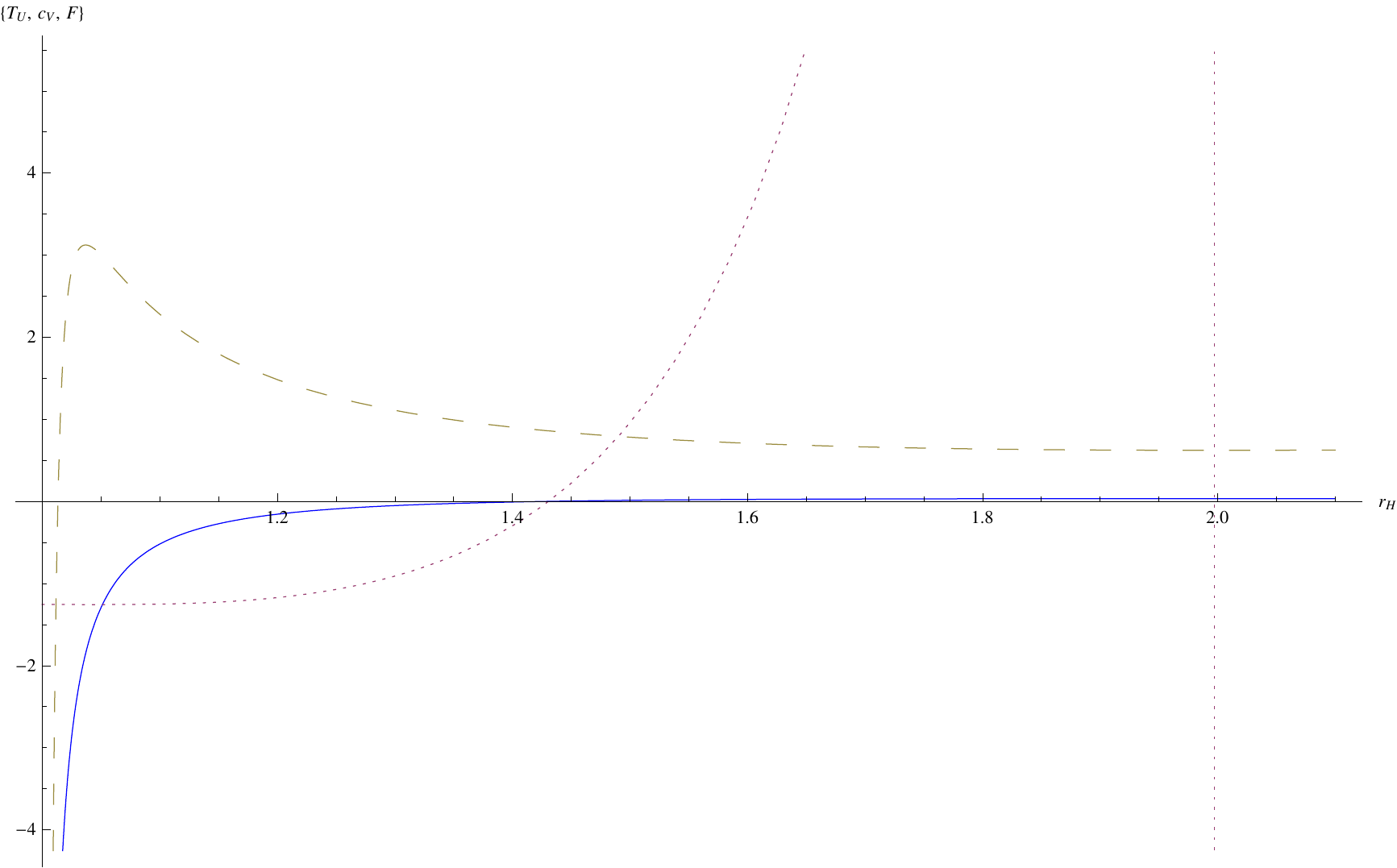}
     }
     \hfill
     \subfloat[$R_{v}=1$ and $2<r_{H}<5$\label{CapacityHeatTemperatureFreeEnergyVectorRange3}]{%
      \includegraphics[width=0.5\textwidth]{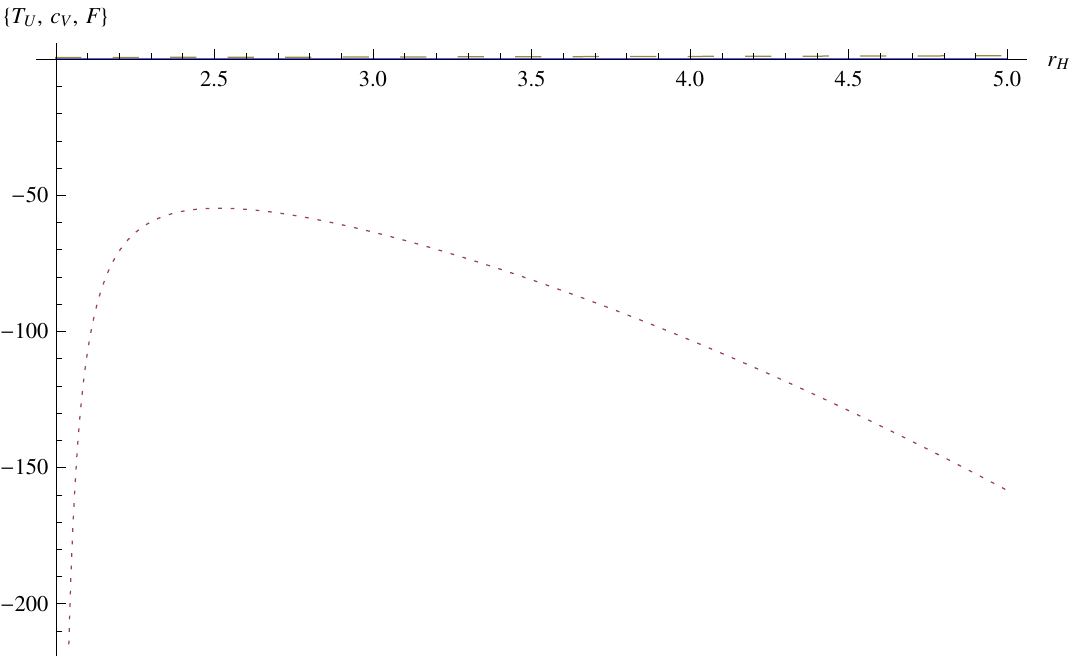}

     }
     \caption{The Un-vector BH heat capacity, temperature and free energy depicted in dotted, solid and dashed lines, respectively.}
    \label{UnvectorThermoVariables}
   \end{figure}

\subsection{Unparticles as Quintessence}
The solution (\ref{staticsolution}) is very similar to the quintessence one \cite{Kiselev:2002dx}, in which
\begin{equation}\label{kiselev}
(g^q_{rr})^{-1}=-g_{00}^q=1-\frac{2M}{r}-\frac{\alpha}{r^{3\omega_q+1}}
\end{equation}
where $\alpha\geq0$ and $\omega$ is constrained by cosmological evidences to $-1\leq\omega\leq-1/3$ \cite{Kiselev:2002dx}. In fact
they become identical if
\begin{equation}\label{mapeamento}
3\omega+1=2d_{U}-1;\; \alpha=\pm2M\Gamma_{U}R_{s,v}^{2d_{U}-2}.
\end{equation}
This suggests that unparticles could be a source of quintessential matter for the static black hole. Since $\omega$ has a range of permitted values, the above relationship between $\omega$ and $d_U$ give us $-1/2\leq d_U\leq 1/2$. However this is not allowed since, by unitarity reasons,   $d_U\geq1$ for scalar (tensor) unparticles and $d_U\geq3$ for the vector ones \cite{Grinstein:2008qk}. Therefore the static black hole quintessence can not be described by unparticles.  Despite of this, the formal similarity of the above solutions will guide us to generalize the unparticle black hole to the case with rotation.

\section{Analytical Solution to the Unparticle Kerr Black Hole}
In this section we construct the solution for the rotating unparticle black hole. The standard procedure is by using the Newman-Janis algorithm. However we can take a shortcut since, as said before, our static solution is formally identical to the one with quintessence found in Ref. \cite{Kiselev:2002dx}. Some time later by using the above algorithm with two different complexifications for the radial coordinate, solutions for the rotating black hole surrounded by quintessence were independently found by Ghosh \cite{Ghosh:2015ovj} and Toshmatov {\it et al.} \cite{Toshmatov:2015npp}. The solutions in the Boyer-Lindquist  coordinates are given by
\begin{align*}
ds^{2} & =\frac{\tilde{\Delta}_{1,2}-a^{2}\sin^{2}\theta}{\Sigma}dt^{2}-\frac{\Sigma}{\tilde{\Delta}_{1,2}}\,dr^{2}+2a\sin^{2}\theta\left(1-\frac{\tilde{\Delta}_{1,2}-a^{2}\sin^{2}\theta}{\Sigma}\right)dt\,d\phi-\Sigma\,d\theta^{2}\\
 & -\,\sin^{2}\theta\left[\Sigma+a^{2}\sin^{2}\theta\left(2-\frac{\tilde{\Delta}_{1,2}-a^{2}\sin^{2}\theta}{\Sigma}\right)\right]d\phi^{2},
\end{align*}
where $\Sigma=r^{2}+a^{2}\cos^{2}\theta$ and
\begin{equation}
\tilde{\Delta}_{1,2}=r^{2}+a^{2}-2Mr-\alpha F_{1,2}^{1-3\omega}.
\end{equation}
The subscripts $1$ ($2$) means the Ghosh (Toshmatov) solutions and the functions $F_{1,2}$ are given by
\begin{equation}
F_1=\Sigma ^\frac{1}{2};\; F_2=r.
\end{equation}

Now we apply our mapping (\ref{mapeamento}) and we get the solution for the rotating unparticle black holes
\begin{equation}\label{unparticlerotantingmetric}
\tilde{\Delta}_U=r^{2}+a^{2}-2Mr\pm (2M)R_{sv}^{2d_{U}-2}F_{1,2}^{3-2d_{U}}.
\end{equation}
From this metric we can obtain the mass parameter and the Hawking temperature of the black holes. The mass parameter $M$ is given by $\tilde{\Delta}=0$ or
\begin{equation}\label{unparticlerotatingmass}
2M=\frac{r_{H}^{2}+a^{2}}{r_{H}\pm R_{sv}^{2d_{U}-2}F_{1,2}^{3-2d_{U}}}.
\end{equation}
The Hawking temperature can also be obtained by using \cite{Wald:1984rg}
\begin{equation}
T_{H}=\frac{1}{4\pi}\frac{\Delta'(r_{H})}{(r_{H}^{2}+a^{2})},
\end{equation}
which provide us the Hawking temperatures
\begin{equation}\label{unparticlerotatingtemperature}
T_{d_{U}}=\frac{1}{2\pi}\left(\frac{r_{H}-M}{r_{H}^{2}+a^{2}}\right)\left[1\pm MR_{s,v}^{2d_{U}-2}\left(\frac{3-2d_{U}}{r_{H}-M}\right)r_{H}F_{1,2}^{1-2d_{U}}\right].
\end{equation}
If in equations (\ref{unparticlerotantingmetric}), (\ref{unparticlerotatingmass}) and (\ref{unparticlerotatingtemperature}) we take the limit $R_{s,v}\to 0$ we see that the standard results for
the Kerr black hole are recovered.

From now on we must perform the analyzes of thermodynamic properties of this solution just in the GD and UGD regimes. This is due to the fact that now the metric depends on the coordinate $\theta$ and we have not been able to find an analytical solution for the entropy. For the temperature we have in the UGD regime
\begin{equation}\label{temperatureUGD}
T_{d_{U}}\approx\frac{1}{4\pi}\left[\frac{2r_{H}}{\left(r_{H}^{2}+a^{2}\right)}+(2d_{U}-3)\frac{r_{H}}{F^{2}_{1,2}}\right],
\end{equation}
where we have considered the more simple case $\theta=\pi/2$. Just as before, if we are looking for a fractalization of the event horizon, we must compare this result with the one for a higher dimensional rotating black hole. This expression is given by \cite{Emparan:2008eg}
\begin{equation}
T_{D}=\frac{1}{4\pi}\left(\frac{2r_{0}}{r_{0}^{2}+a^{2}}+\frac{D-5}{r_{0}}\right).
\end{equation}
Comparing this with our UGD expression (\ref{temperatureUGD}) we get a fractalization of the event horizon in the Ghosh-like case is given by
\begin{equation} \label{rotatingfractalization}
d_H=3+\frac{2d_U-3}{1+\left(\frac{a}{r_H}\right)^2\cos^2\theta}
\end{equation}
and for the Toshmatov-like solution we get $d_H=2d_U$ as in the static unparticle black hole case.
We should point out that these solutions reduce to case without rotation if we take $a=0$.

\section{Conclusions}
In this paper we initially perform a detailed study of the thermodinamic properties of scalar (tensor) and vector unparticle static black holes.  For this we have found analytical expressions for the main thermodynamic quantities: Hawking temperature, entropy, heat capacity and free energy. For the scalar (tensor) cases we first analyzed the Hawking temperature and found a behavior very similar to the case without unparticles, as can be seem in Fig. \ref{TempUngravity}. The same trivial behavior is found for heat capacity and free energy. However, when one studies the entropy, it is obtained that black holes have a remnant minimum value for the entropy when its radius approaches zero, what is diverse from the conventional case. This conclusion emerges directly from our analytical solution (\ref{scalarentropy}) depicted in Fig. \ref{EntropyUngravity}. We have found that this is the only possible solution that recover standard black holes in the limit without unparticles ($R_s \to 0$).

For the vector case we found a more involved structure. In this case all the analyzed quantities are singular at $r_H=R_v$.
This suggests that we cannot transit from one region to the other and therefore it seems that we have disjoined regions $r_H\gtrless R_v$.
First of all the mass parameter is singular at $r_H=R_v$ as can be seen from Eq. (\ref{statichorizon}). This equation also points to the fact that the mass parameter is always negative in the domain $r_H<R_v$. Despite of this apparently forbidden region of  negative mass, recent works points to the fact that black holes with $M<0$ can exist and present non-trivial topology \cite{Mann:1997jb}. Since the ungravity dominated regime must be contained in a full quantum gravity theory, it is expected that
quantum fluctuations in geometry should enhance quantum fluctuations in topology \cite{Wheeler:1957mu}. Therefore we put forward the study of other thermodynamic quantities. For example, when we investigated the Hawking temperature, we found a minimum value $T_0$ at the point $r^{-}_H$ (see Eq. \ref{roots}) exactly in the above pointed region as can be seem in Fig. \ref{Un-vectorThermoVariables}. This suggests the existence of a phase transition. In order to confirm this, we computed the heat capacity and the free energy.  By analyzing the former we identified that it is singular at the point where the temperature is a local minimum and it is positive (negative) if $r_H<r^{-}_H$ ($r_H>r^{-}_H$). This implies that the vector unparticle black hole is thermodynamically stable (unstable) in these regions. Regarding free energy we found that the constant of integration $A_v$ had to be fixed in order to be consistent with a phase transition at $r_H=r^{-}_H$. This means that at this point the free energy is null and therefore the global phase transition corresponds to the local one. Therefore as well as in the scalar (tensor) cases we also get a remnant value for the entropy if the black hole evaporate completely. This can be seen in Fig. \ref{EntropyUngravity}.

In another direction, we have shown that unparticles cannot be a source of quintessence. For this we first point that the static unparticle solution found in Refs.  \cite{Gaete:2010sp,Mureika:2010je} and the quintessence one found by Kiselev in Ref. \cite{Kiselev:2002dx} are formally similar. They in fact become identical if the mapping (\ref{mapeamento}) is considered. With this we get that the constraints in the scale dimension $d_U$ ($d_U>2$) is not compatible with the range of the observational quintessential parameter $-1<\omega<-1/3$.

The above identification also pointed to us a shortcut to obtain the rotating solutions of the systems under consideration. By using two different complexifications in the Newman-Janis algorithm, Ghosh and Toshmatov {\it et al} found two distinct solutions to the rotating black hole surrounded by quintessence.  By applying (\ref{mapeamento}) we found, for the first time, the corresponding exact solutions to the rotating unparticle black holes. For both cases we computed the Hawking temperature.  In the ungravity dominated regime we found, as in the static cases, a fractalization of the event horizon. For the Toshmatov-like one it is equal to the static case and therefore the fractalization is not dependent on the rotation of the source. However for the Gosh-like solution the fractal dimension depends on the polar angle and on the rotation of the source.

\subsection*{Acknowledgments}
The authors would like to thank Alexandra Elbakyan, for removing all barriers in the way of science.
We acknowledge the financial support provided by the Conselho
Nacional de Desenvolvimento Cient\'\i fico e Tecnol\'ogico (CNPq) and Funda\c c\~ao Cearense de
Apoio ao Desenvolvimento Cient\'\i fico e Tecnol\'ogico (FUNCAP) through PRONEM PNE-0112-00085.01.00/16.


\begin{thebibliography}{10}

\bibitem{ZinnJustin:2002ru}
  J.~Zinn-Justin,
  Int.\ Ser.\ Monogr.\ Phys.\  {\bf 113}, 1 (2002).

\bibitem{Polchinski:1987dy}
  J.~Polchinski,
  Nucl.\ Phys.\ B {\bf 303}, 226 (1988).
  doi:10.1016/0550-3213(88)90179-4

\bibitem{Horava:2009uw}
  P.~Horava,
  Phys.\ Rev.\ D {\bf 79}, 084008 (2009)
  doi:10.1103/PhysRevD.79.084008
  [arXiv:0901.3775 [hep-th]].

\bibitem{Horava:2008ih}
  P.~Horava,
  JHEP {\bf 0903}, 020 (2009)
  doi:10.1088/1126-6708/2009/03/020
  [arXiv:0812.4287 [hep-th]].

\bibitem{Horava:2009if}
  P.~Horava,
  Phys.\ Rev.\ Lett.\  {\bf 102}, 161301 (2009)
  doi:10.1103/PhysRevLett.102.161301
  [arXiv:0902.3657 [hep-th]].

\bibitem{Visser:2011mf}
  M.~Visser,
  J.\ Phys.\ Conf.\ Ser.\  {\bf 314}, 012002 (2011)
  doi:10.1088/1742-6596/314/1/012002
  [arXiv:1103.5587 [hep-th]].

\bibitem{Kehagias:2009is}
  A.~Kehagias and K.~Sfetsos,
  Phys.\ Lett.\ B {\bf 678}, 123 (2009)
  doi:10.1016/j.physletb.2009.06.019
  [arXiv:0905.0477 [hep-th]].

\bibitem{Muniz:2013uva}
  C.~R.~Muniz, V.~B.~Bezerra and M.~S.~Cunha,
  Phys.\ Rev.\ D {\bf 88}, 104035 (2013)
  doi:10.1103/PhysRevD.88.104035
  [arXiv:1311.2570 [gr-qc]].

\bibitem{Muniz:2014dga}
  C.~R.~Muniz, V.~B.~Bezerra and M.~S.~Cunha,
  Annals Phys.\  {\bf 359}, 55 (2015)
  doi:10.1016/j.aop.2015.04.014
  [arXiv:1405.5424 [hep-th]].

\bibitem{Alencar:2015aea}
  G.~Alencar, V.~B.~Bezerra, M.~S.~Cunha and C.~R.~Muniz,
  Phys.\ Lett.\ B {\bf 747}, 536 (2015)
  doi:10.1016/j.physletb.2015.06.049
  [arXiv:1505.05087 [hep-th]].

\bibitem{Banks:1981nn}
  T.~Banks and A.~Zaks,
  Nucl.\ Phys.\ B {\bf 196}, 189 (1982).
  doi:10.1016/0550-3213(82)90035-9

\bibitem{Georgi:2007ek}
  H.~Georgi,
  Phys.\ Rev.\ Lett.\  {\bf 98}, 221601 (2007)
  doi:10.1103/PhysRevLett.98.221601
  [hep-ph/0703260].

\bibitem{Georgi:2007si}
  H.~Georgi,
  Phys.\ Lett.\ B {\bf 650}, 275 (2007)
  doi:10.1016/j.physletb.2007.05.037
  [arXiv:0704.2457 [hep-ph]].

\bibitem{Luo:2007bq}
  M.~Luo and G.~Zhu,
  Phys.\ Lett.\ B {\bf 659}, 341 (2008)
  doi:10.1016/j.physletb.2007.10.058
  [arXiv:0704.3532 [hep-ph]].

\bibitem{Aliev:2007qw}
  T.~M.~Aliev, A.~S.~Cornell and N.~Gaur,
  Phys.\ Lett.\ B {\bf 657}, 77 (2007)
  doi:10.1016/j.physletb.2007.09.055
  [arXiv:0705.1326 [hep-ph]].

\bibitem{Chen:2007qr}
  S.~L.~Chen and X.~G.~He,
  Phys.\ Rev.\ D {\bf 76}, 091702 (2007)
  doi:10.1103/PhysRevD.76.091702
  [arXiv:0705.3946 [hep-ph]].

\bibitem{Mathews:2007hr}
  P.~Mathews and V.~Ravindran,
  Phys.\ Lett.\ B {\bf 657}, 198 (2007)
  doi:10.1016/j.physletb.2007.10.018
  [arXiv:0705.4599 [hep-ph]].

\bibitem{Bander:2007nd}
  M.~Bander, J.~L.~Feng, A.~Rajaraman and Y.~Shirman,
  Phys.\ Rev.\ D {\bf 76}, 115002 (2007)
  doi:10.1103/PhysRevD.76.115002
  [arXiv:0706.2677 [hep-ph]].

\bibitem{Cheung:2007ap}
  K.~Cheung, W.~Y.~Keung and T.~C.~Yuan,
  Phys.\ Rev.\ D {\bf 76}, 055003 (2007)
  doi:10.1103/PhysRevD.76.055003
  [arXiv:0706.3155 [hep-ph]].

\bibitem{Kikuchi:2007qd}
  T.~Kikuchi and N.~Okada,
  Phys.\ Lett.\ B {\bf 661}, 360 (2008)
  doi:10.1016/j.physletb.2008.02.041
  [arXiv:0707.0893 [hep-ph]].

\bibitem{Zhang:2007ih}
  H.~Zhang, C.~S.~Li and Z.~Li,
  Phys.\ Rev.\ D {\bf 76}, 116003 (2007)
  doi:10.1103/PhysRevD.76.116003
  [arXiv:0707.2132 [hep-ph]].

\bibitem{Cheung:2007jb}
  K.~Cheung, W.~Y.~Keung and T.~C.~Yuan,
  In *Karlsruhe 2007, SUSY 2007* 698-701
  [arXiv:0710.2230 [hep-ph]].




\bibitem{Lewis:2007ss}
  I.~Lewis,
  arXiv:0710.4147 [hep-ph].

\bibitem{McDonald:2007bt}
  J.~McDonald,
  JCAP {\bf 0903}, 019 (2009)
  doi:10.1088/1475-7516/2009/03/019
  [arXiv:0709.2350 [hep-ph]].

\bibitem{Davoudiasl:2007jr}
  H.~Davoudiasl,
  Phys.\ Rev.\ Lett.\  {\bf 99}, 141301 (2007)
  doi:10.1103/PhysRevLett.99.141301
  [arXiv:0705.3636 [hep-ph]].

\bibitem{Das:2007nu}
  P.~K.~Das,
  Phys.\ Rev.\ D {\bf 76}, 123012 (2007)
  doi:10.1103/PhysRevD.76.123012
  [arXiv:0708.2812 [hep-ph]].

\bibitem{Hannestad:2007ys}
  S.~Hannestad, G.~Raffelt and Y.~Y.~Y.~Wong,
  Phys.\ Rev.\ D {\bf 76}, 121701 (2007)
  doi:10.1103/PhysRevD.76.121701
  [arXiv:0708.1404 [hep-ph]].

\bibitem{Alberghi:2007vc}
  G.~L.~Alberghi, A.~Y.~Kamenshchik, A.~Tronconi, G.~P.~Vacca and G.~Venturi,
  Phys.\ Lett.\ B {\bf 662}, 66 (2008)
  doi:10.1016/j.physletb.2008.01.076
  [arXiv:0710.4275 [hep-th]].



\bibitem{Gaete:2010sp}
  P.~Gaete, J.~A.~Helayel-Neto and E.~Spallucci,
  Phys.\ Lett.\ B {\bf 693}, 155 (2010)
  doi:10.1016/j.physletb.2010.07.058
  [arXiv:1005.0234 [hep-ph]].

\bibitem{Mureika:2010je}
  J.~R.~Mureika and E.~Spallucci,
  Phys.\ Lett.\ B {\bf 693}, 129 (2010)
  doi:10.1016/j.physletb.2010.08.025
  [arXiv:1006.4556 [hep-ph]].

\bibitem{Mureika:2007nc}
  J.~R.~Mureika,
  Phys.\ Lett.\ B {\bf 660}, 561 (2008)
  doi:10.1016/j.physletb.2008.01.050
  [arXiv:0712.1786 [hep-ph]].

\bibitem{Frassino:2013lya}
  A.~M.~Frassino, P.~Nicolini and O.~Panella,
  Phys.\ Lett.\ B {\bf 772}, 675 (2017)
  doi:10.1016/j.physletb.2017.07.029
  [arXiv:1311.7173 [hep-ph]].


\bibitem{Kiselev:2002dx}
  V.~V.~Kiselev,
  Class.\ Quant.\ Grav.\  {\bf 20}, 1187 (2003)
  doi:10.1088/0264-9381/20/6/310
  [gr-qc/0210040].








\bibitem{Mann:1997jb}
  R.~B.~Mann,
  Class.\ Quant.\ Grav.\  {\bf 14}, 2927 (1997)
  doi:10.1088/0264-9381/14/10/018
  [gr-qc/9705007].

\bibitem{Wheeler:1957mu}
  J.~A.~Wheeler,
  Annals Phys.\  {\bf 2}, 604 (1957).
  doi:10.1016/0003-4916(57)90050-7

\bibitem{Wald:1984rg}
  R.~M.~Wald,
  doi:10.7208/chicago/9780226870373.001.0001

\bibitem{Emparan:2008eg}
  R.~Emparan and H.~S.~Reall,
  Living Rev.\ Rel.\  {\bf 11}, 6 (2008)
  doi:10.12942/lrr-2008-6
  [arXiv:0801.3471 [hep-th]].



\bibitem{Grinstein:2008qk}
  B.~Grinstein, K.~A.~Intriligator and I.~Z.~Rothstein,
  Phys.\ Lett.\ B {\bf 662}, 367 (2008)
  doi:10.1016/j.physletb.2008.03.020
  [arXiv:0801.1140 [hep-ph]].

\bibitem{Ghosh:2015ovj}
  S.~G.~Ghosh,
  Eur.\ Phys.\ J.\ C {\bf 76}, no. 4, 222 (2016)
  doi:10.1140/epjc/s10052-016-4051-7
  [arXiv:1512.05476 [gr-qc]].

\bibitem{Toshmatov:2015npp}
  B.~Toshmatov, Z.~Stuchlík and B.~Ahmedov,
  Eur.\ Phys.\ J.\ Plus {\bf 132}, no. 2, 98 (2017)
  doi:10.1140/epjp/i2017-11373-4
  [arXiv:1512.01498 [gr-qc]].



\end{thebibliography}
\end{document}